# Automatic Blink-based Bad EEG channels Detection for BCI Applications*

Eva Guttmann-Flury[1], Yanyan Wei[2], and Shan Zhao[3]

*Abstract*— In Brain-Computer Interface (BCI) applications, noise presents a persistent challenge, often compromising the quality of EEG signals essential for accurate data interpretation. This paper focuses on optimizing the signal-to-noise ratio (SNR) to improve BCI performance, with channel selection being a key method for achieving this enhancement. The Eye-BCI multimodal dataset is used to address the issue of detecting and eliminating faulty EEG channels caused by non-biological artifacts, such as malfunctioning electrodes and power line interference. The core of this research is the automatic detection of problematic channels through the Adaptive Blink-Correction and De-Drifting (ABCD) algorithm. This method utilizes blink propagation patterns to identify channels affected by artifacts or malfunctions. Additionally, segmented SNR topographies and source localization plots are employed to illustrate the impact of channel removal by comparing Left and Right hand grasp Motor Imagery (MI). Classification accuracy further supports the value of the ABCD algorithm, reaching an average classification accuracy of 93.81% [74.81%; 98.76%] (confidence interval at 95% confidence level) across 31 subjects (63 sessions), significantly surpassing traditional methods such as Independent Component Analysis (ICA) (79.29% [57.41%; 92.89%]) and Artifact Subspace Reconstruction (ASR) (84.05% [62.88%; 95.31%]). These results underscore the critical role of channel selection and the potential of using blink patterns for detecting bad EEG channels, offering valuable insights for improving real-time or offline BCI systems by reducing noise and enhancing signal quality.

## I. INTRODUCTION

Noise is an inherent challenge in signal processing, particularly in Brain-Computer Interface (BCI) applications where the signal of interest, originating from brain activity, is often overwhelmed by various interferences. Both biological and non-biological artifacts significantly degrade the signal-to-noise ratio (SNR), making reliable detection of brain patterns particularly demanding in practical BCI systems.

Recording equipment failures and environmental interference (e.g., power line noise) systematically corrupt EEG signals with non-biological artifacts, creating persistent barriers to reliable data interpretation. Selective removal of affected channels serves as a critical preprocessing step, directly enhancing the reliability of subsequent BCI signal analysis and classification.

Channel selection algorithms facilitate the identification and exclusion of noisy or irrelevant channels, which in turn enhances system performance by reducing overfitting and computational complexity [1]. Furthermore, reducing the number of EEG channels by eliminating compromised ones renders BCI systems more user-friendly and practical for real-world applications. This reduction can simplify the setup process, minimize user discomfort, and lower costs. Optimizing the number of channels without sacrificing system performance is critical for the practical deployment of BCI frameworks in everyday life [2] [3].

Eliminating poor-quality channels is crucial for maintaining signal integrity and ensuring reliable BCI performance. Degraded channels introduce artifacts that compromise accuracy and robustness. Advanced algorithms have been developed to address this, including statistical methods like the Local Outlier Factor (LOF) [4] and Iterative Standard Deviation Method [5], which identify noisy channels with high accuracy. Machine learning techniques, such as ensemble bagging classifiers [6] and deep convolutional neural networks (CNNs) [7], automate channel classification and reduce manual feature extraction. Additionally, attention mechanisms [8] and graph theory-based approaches [9] enhance reliability by weighting channels or modeling the brain as a network to detect faulty channels.

Traditional methods for bad channel detection typically rely on statistical analyses or machine-learning algorithms. In contrast, the current approach explores an alternative strategy by leveraging natural physiological properties of blink propagation (occurring approximately 20 times per minute) to reliably identify faulty electrodes. Implementing periodic quality checks (e.g., every 5 minutes) could ensure sufficient blink data for robust channel assessment while preserving system responsiveness. Further, this near-real-time framework might be extended to include adaptive triggering based on performance metrics, allowing for verification when classification accuracy declines. Such a combination of physiological principles and adaptive monitoring could offer a practical solution for sustaining signal quality and enhancing classification accuracy in BCI systems.

*This work is supported in part by STI 2030 Major Projects, the National Key Research and Development Program of China and the Natural Science Foundation of China (Grant No. 2022ZD0208500).

[1]Eva Guttmann-Flury is with the Department of Micro-Nano Electronics and the MoE Key Laboratory of Artificial Intelligence, Shanghai Jiao Tong University, Shanghai, China `eva.guttmann.flury@gmail.com`

[2]Yanyan Wei is with the Shanghai Key Laboratory of Psychotic Disorders Shanghai Mental Health Center, Shanghai Jiao Tong University School of Medicine, Shanghai, China `weiyanyan19860729@126.com`

[3]Shan Zhao is with the School of Public Health, Shanghai Jiao Tong University School of Medicine, Shanghai, China `shanzhao23@sjtu.edu.cn`

## II. METHODS

Building on blink propagation physiology, this work develops and validates an end-to-end pipeline for bad EEG channel detection. The subsequent sections detail the experimental parameters, participant cohort, SNR quantification, and adaptive algorithms that collectively enable reliable signal quality monitoring in BCIs.

### A. Study design

This study analyzed the Eye-BCI multimodal dataset, hosted on Synapse [10], containing simultaneous 62-channel EEG signals, electrooculogram (EOG), electromyogram (EMG), high-speed video, and synchronized eye-tracking. Four paradigms were recorded: Motor Imagery (MI), Motor Execution (ME), Steady-State Visually Evoked Potentials (SSVEP) (each containing 2,520 trials), and P300 spellers (with 5,670 trials), totaling over 46 hours of multimodal data [11]. These paradigms were selected to elicit various sensory-motor responses and assess their impact on ocular activity. Experimental setup instructions and code for replicating the dataset are available at https://github.com/QinXinlan/EEG-experiment-to-understand-differences-in-blinking/.

### B. Objectives

This work's primary goal is to develop and validate a novel method for automatically detecting faulty EEG channels by leveraging the physiological characteristics of blink propagation. This approach hypothesizes that the consistent spatiotemporal characteristics of blink patterns can be leveraged to achieve highly accurate detection with minimal computational demand, making it suitable for real-time or near-real-time processing.

### C. A priori sample size

The phenomenon of interest is here the spontaneous blink within EEG signals. Defined as the maximum potential value when the upper eyelid approaches the lower eyelid, the peak amplitude is identified as the variable of interest. Given that the distribution of this parameter is non-Gaussian, a Fitted Distribution Monte Carlo (FDMC) simulation was conducted. Results indicated that a minimum of 45 sessions is required to ensure an adequate sample size, assuming an effect size evaluated with a conservative Cohen's d = 0.2 [12].

### D. Participants

This study was approved by the Institutional Review Board of Shanghai Jiao Tong University, Protocol No. (IRB HRP E2021216I), date of approval (March 4th, 2021), and all participants provided informed consent, permitting the collection and publication of anonymized data. Thirty-one healthy adults (11 women and 20 men, aged 20 to 57) volunteered for the study. Fourteen participants completed a single session, two attended two sessions, and fifteen completed three sessions, totaling 63 sessions overall. To prevent voluntary blinking, participants were briefed on the BCI paradigm goals without reference to blinks.

### E. Signal-to-Noise Ratio (SNR)

The signal-to-noise ratio (SNR) is the dimensionless ratio of the desired signal power to the background noise power. It can be measured for a single trial ($SNR_T$) or the grand average (GA) of all trials ($SNR_{GA}$). The primary goal of computing the GA is to attenuate the noise by averaging across multiple trials, thereby making the signal stand out more prominently. This process typically results in higher signal power and lower noise, leading to an improved SNR. When brain patterns are obscured by background EEG, the SNR is small, making detection difficult. A higher SNR, however, improves BCI detection and classification. Enhancing the SNR is critical for reliably estimating single-trial responses and also allows for minimizing the sample size required to achieve significant results. The measured signal for a single trial can be represented as the sum of the mean signal of interest (SOI) and noise, with the SOI assumed constant and independent of noise, and noise assumed stationary and ergodic. The power of the measured signal is the expected value of its squared value.

$$\mathbb{P}(x_i) = \mathbb{E}[x_i^2] = \mathbb{E}[s_i^2 + 2 * s_i * n_i^b + (n_i^b)^2]$$
$$= \mathbb{E}[s_i^2] + 2 * \mathbb{E}[s_i] * \mathbb{E}[n_i^b] + \mathbb{E}[(n_i^b)^2] \quad (1)$$

Since the mean of the basic noise is assumed to be equal to zero, $\mathbb{E}[n_i^b] = \overline{n^b} = 0$, the power of the measured signal is simply:

$$\mathbb{P}(x_i) = \mathbb{E}[s_i^2] + \mathbb{E}[(n_i^b)^2] = \mathbb{P}(s_i) + \mathbb{P}(n_i^b) \quad (2)$$

The SNR of a single trial can then be expressed as the ratio between the signal power $\mathbb{P}(s_i)$ and the basic noise power $\mathbb{P}(n_i^b)$. The centered basic noise $n^b$ is assumed stationary and ergodic, so its power can be calculated over any time interval. Therefore, the noise power can be directly computed from the recorded signal before (or after) the SOI. Let $t_n$ (resp. $t_s$) denote the start time of the noise (resp. SOI), and $\Delta t_n$ (resp. $\Delta t_s$) represent the duration of the noise (resp. SOI) interval. The SNR can then be calculated as follows:

$$SNR_T = \frac{\frac{1}{N}\sum_{i=1}^{N}\frac{1}{\Delta t_s}\int_{t_s}^{t_s+\Delta t_s} x_i^2(t)}{\frac{1}{N}\sum_{i=1}^{N}\frac{1}{\Delta t_n}\int_{t_n}^{t_n+\Delta t_n} x_i^2(t)} - 1 \quad (3)$$

This SNR formulation provides a quantitative framework for assessing signal quality in EEG analysis. However, reliable SNR estimation requires artifact-free recordings — a challenge addressed by the following preprocessing strategy.

## F. The Adaptive Blink Correction and De-drifting (ABCD) algorithm

The ABCD algorithm is designed for real-time applications, prioritizing rapid detection with minimal computational complexity [13]. To achieve this, the detection step combines amplitude and attenuation criteria, leveraging the way blinks propagate across the scalp.

*1) Blink Detection:* The blink detection process begins by preprocessing a frontopolar channel (e.g., FP1) data (200Hz downsampling, 10Hz low-pass filtering) in 2-second windows. Potential blinks are first identified using an amplitude threshold compared to a moving average. Candidate events then undergo rigorous shape verification: the pre-amplitude check analyzes the 200 ms window preceding the peak, while the post-amplitude check evaluates the subsequent 300 ms interval. Spatial validation is then performed via attenuation comparison between a frontopolar (e.g., FP1) and a central (e.g., CZ) electrode. Only events satisfying all three criteria (amplitude, shape, and propagation) are classified as valid blinks, while artifacts and noise are systematically rejected. This multi-stage filtering ensures robust detection while efficiently discarding blink-free segments early in the process, as illustrated in Figure 1.

*2) Blink Propagation:* In a simplified blink propagation model, the eyes can be seen as point sources, with the blink generating a single wave (or pulse) that propagates through an anisotropic medium. Since EEG electrodes are positioned on the skull, this propagation can be approximated as waves traveling across a 2D sphere (the head). The electric field generated by a blink creates a potential following electrostatic principles, resulting in an attenuation factor of $\frac{1}{r^2}$ relative to an equivalent dipole centered at the FP1 reference electrode:

$$V_{th} = \frac{Q}{4\pi\epsilon_r d_{FP1}^2} \quad (4)$$

Where $Q$ denotes the unknown dipole moment and $d$ the distance to the dipole's center, located at FP1. If the synthetic permittivity value were available, the dipole moment's projected norm could be calculated using normalized distances within the framework of these approximations. Conversely, given a known artificial dipole, this synthetic permittivity could be deduced.

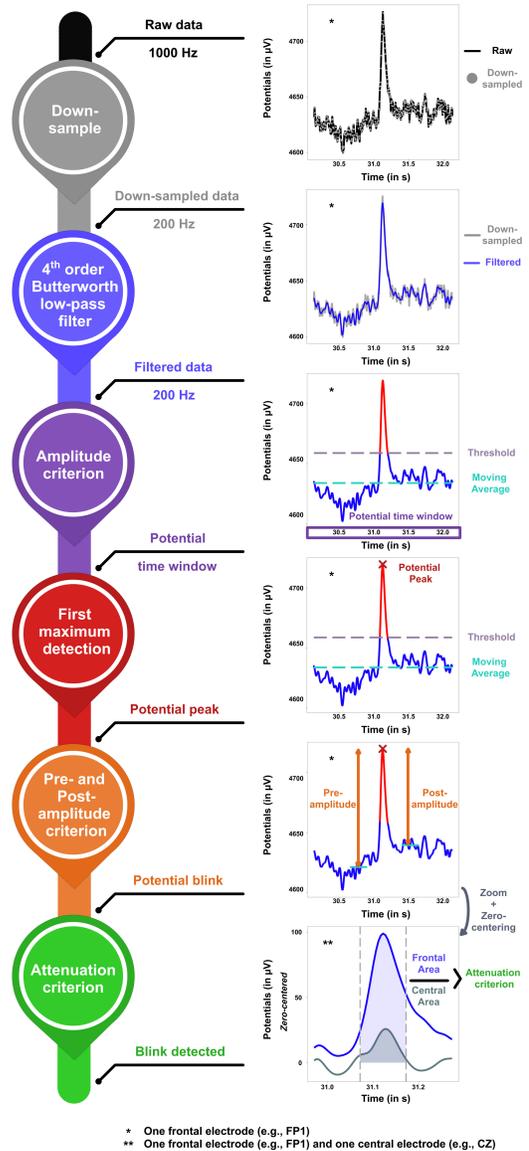

Fig. 1. Blink detection flowchart in the Adaptive Blink Correction and De-drifting (ABCD) algorithm

dipole-centric distances ($d_{FP1}$) are computed from inter-electrode measurements in the MNI Average Brain (305 MRI) Stereotaxic Registration Model [15].

TABLE I
Experimental and theoretical potential propagation comparison at central electrodes

| Electrodes | FZ | FCZ | CZ | CPZ | PZ |
|---|---|---|---|---|---|
| $d_{FP1}$ | 86.12 | 118.00 | 148.33 | 173.80 | 192.28 |
| $V$ | 54.35 | 31.11 | 21.22 | 16.35 | 13.83 |
| $V_{th}$ | 57.92 | 32.32 | 21.38 | 16.25 | 13.83 |
| $w_s$ | 0.84 | 0.88 | 0.92 | 0.96 | 1 |

The skull thickness ratio values ($w_s$) in Table I are linearly scaled according to established neuroanatomical data [16], accounting for regional variations in cranial bone density between sexes (Table II). Concurrently, the

TABLE II
Gender differences in typical skull thickness

| Gender | Female | Male |
|---|---|---|
| Frontal skull thickness (in mm) | 8.6 | 7.8 |
| Parietal skull thickness (in mm) | 10.1 | 9.6 |
| Frontal $w_s$ (ratio at FZ) | 0.85 | 0.81 |

The measured potentials ($V$) represent empirical grand averages computed across all subjects in the current dataset. By setting the skull thickness ratio ($w_s$) to 1 at the farthest electrode, the equality between the measured ($V$) and theoretical ($V_{th}$) potentials allows for the direct computation of the system's constant ($\frac{Q}{4\pi\epsilon_r}$).

Using this constant, the theoretical potential at each electrode, summarized in Table I, can be computed with the following attenuation model, adjusted for skull thickness:

$$V_{th} = \frac{Qw_s}{4\pi\epsilon_r d_{FP1}^2} \quad (5)$$

The close agreement between these predictions and empirical measurements validates the model's biophysical foundations. Figures 2 and 3 both illustrate the comparison between theoretical and observed potentials, with Figure 2 providing a clearer depiction of the near-perfect alignment at the midline electrodes.

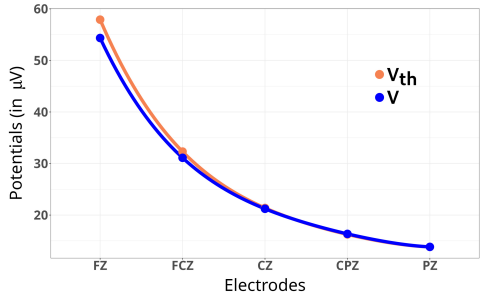

Fig. 2. Quasi-perfect match of theoretical and real potential (from grand averages) attenuated propagation at the central electrodes

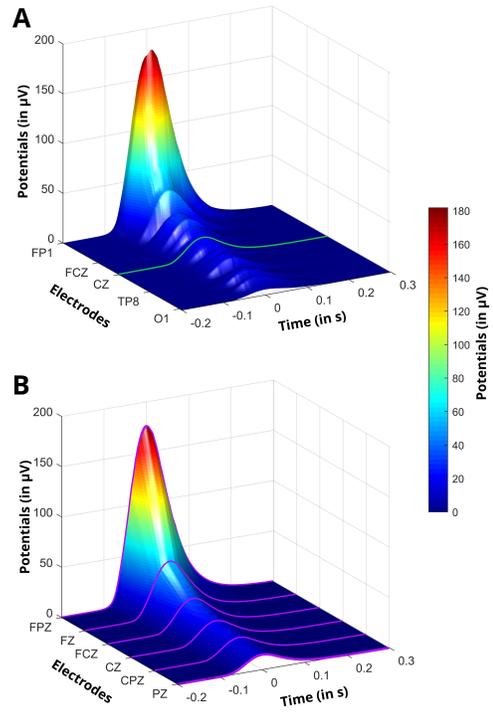

Fig. 3. (A): Real grand average blink propagation across all electrodes (the electrode used for the propagation criterion - CZ - is in green); (B): Theoretical grand average blink propagation (midline electrodes are in magenta)

As the distance from the eyes increases, the blink amplitude is expected to diminish. The overall waveform recorded by more distant electrodes retains a similar shape, although the blink's base width (analogous to a wavelength in a sinusoidal wave) also decreases with distance.

Figure 3A displays the real grand average blink potentials across all electrodes for a representative subject during ∼5 minutes of recording, with the central channel, CZ, highlighted as the reference electrode for ABCD's attenuation criterion. Figure 3B illustrates the theoretical grand average blink potentials, computed using the real values from the frontal reference electrode, FP1, and the theoretical attenuation coefficients derived from equation 5, showing the characteristic monotonic amplitude decrease along midline electrodes.

*3) Bad Channel Removal:* EEG channels may produce improper recordings due to technical issues such as abnormal impedance, broken wire contacts, or excessive conductive gel forming bridges between electrodes. When a channel malfunctions, its signals often differ significantly from those of neighboring functional channels, and in severe cases, may even appear as a flat line. In contrast, bridged electrodes yield identical signals across affected channels.

To automatically detect these issues, blink activity can be analyzed. The Blink-Related Potential (BRP) is computed for each electrode and compared to the median BRP of its neighboring electrodes, which minimizes the impact of outliers. The longest common subsequence (LCSS) between these two BRP trajectories is calculated; if the LCSS is below a defined threshold, the electrode is labeled as malfunctioning or "bad" (see Figure 4).

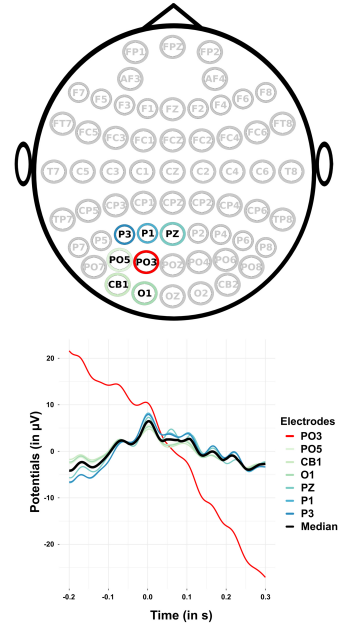

Fig. 4. Identification of bad channels using the blink-related potential (BRP) on neighboring electrodes

In cases of bridged electrodes, the BRPs on affected channels are indistinguishable, and a threshold on the BRP time-series difference is applied to reliably identify bridged or "bad" electrodes.

Let $e_i$ represent the electrode on which the $BRP_i$ is calculated, and $e_{i,n}$ denote its neighboring electrodes. The process for detecting bad channels is summarized in Algorithm 1.

**Data:** the blink-related potential and $e_{i,n}$ at the neighboring electrodes, $L$ threshold for LCSS bad channel detection
**Result:** bad and bridged channel
Initialize bad and bridged electrodes:
$bad = []; bridged = [];$
**for** $i = 1$ *to* $E$
// Loop over all $E$ electrodes
**do**
$BRP_{i,n} = \text{median}(BRP_j)$ with $e_j \in e_{i,n}$
// BRP at the neighboring electrodes
**if** $LCSS(BRP_i, BRP_{i,n}) = 0$ **then**
$bridge = [bridge, i]$
// Bridged electrode detected
**else if** $LCSS(BRP_i, BRP_{i,n}) < L$ **then**
$bad = [bad, i]$
// Bad electrode detected
**end**
**return** $(bad, bridge)$

**Algorithm 1:** Pseudo-code for bad channel detection

*4) Drift Curve Removal:* Low-frequency signal drift, typically caused by sweat or drying conductive gel, constitutes another critical non-biological artifact that biases electrode potentials. The ABCD algorithm addresses this by computing drift curves for each channel using a 4th-order 2 Hz low-pass Butterworth filter. To preserve blink morphology, the drift estimate is replaced with spline interpolation ($D_{SB} = Spline[B_4(s_B)]$) in blink-affected windows, yielding corrected signals ($s_B - D_{SB}$). This order of operations — drift removal following bad channel detection — was empirically determined to optimize computational efficiency while ensuring artifact-free signals for subsequent analysis.

*5) Timing Considerations and Recommendations:* The temporal resolution of the ABCD algorithm is physiologically constrained by blink characteristics. Individual blink validation requires a fixed 500 ms analysis window (200 ms pre-peak + 300 ms post-peak), but reliable bad channel detection necessitates grand averages across multiple blinks due to substantial intra- and inter-subject variability. Across the cohort (N=31), blink peak potentials averaged 160.1 ± 56.4 $\mu$V (mean ± SD) with a frequency of 20.8 ± 12.8 blinks/minute [11]. These parameters motivate our recommendation for periodic quality assessments using ∼5-minute data segments. Adaptive implementations may alternatively employ moving windows of equivalent duration, particularly useful when classification performance declines unexpectedly, enabling continuous verification of potential electrode failures.

The analysis pipeline standardized electrode selection by excluding channels exhibiting malfunctions in any recording session of the Eye-Blink multimodal dataset. From the original 62-electrode montage, 11 channels were removed: 10 due to persistent malfunctions (PO3, F1, POZ, OZ, F3, O2, P8, PO7, FC3, P7) and P4 due to bridging effects. This yielded a consistent 51-electrode setup across all subjects, minimizing bias from inter-session hardware variability.

## III. RESULTS

Large-scale patterns of synchronized neuronal activity exhibit inherent nonlinearity and nonstationarity, with mean and covariance properties that typically shift across different time segments during specific tasks. This non-stationarity is assumed to reflect stages within a self-organized process [17].

### A. Segmented SNR topographies

The variability across segments or trials, along with time-varying relationships between measurement channels, contributes to the non-Gaussian nature of the signals. Extensive studies have documented Event-Related Desynchronization (ERD) in the beta band during Motor Imagery (MI), seen as a reduction in power.

After removal of bad channels and identification of the signal of interest — derived from repetitive patterns computed from source localization methods, though this is not the focus of this paper — the EEG data from the Eye-BCI multimodal dataset are averaged per subject, and SNR plots are computed as per Equation (3). The topographical SNR distribution for beta activity (12.5-30 Hz) during Left and Right MI is illustrated in Figure 5 for a specific subject.

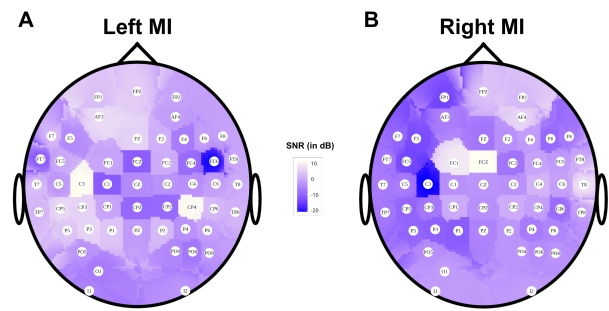

Fig. 5. Segmented SNR topographies in the beta band (12.5-30 Hz) for both Left and Right MI (only the "good" electrodes are represented)

The SNR plots represent the time-averaged values at each electrode. Source localization, using eLORETA, can also be computed for each time point and displayed at the time corresponding to the Motor Imagery task. A 3D eLORETA representation in the beta band for both Left and Right MI is shown in Figure 6.

The comparison of these two plot types highlights the consistency, whether deliberately chosen or emerging from the computations, typically seen in most conventional plots. However, we argue that the inherent irregularity of the recorded EEG signals should be reflected in the representation. This approach does not affect subsequent analyses but emphasizes the unique characteristics of the signals being studied.

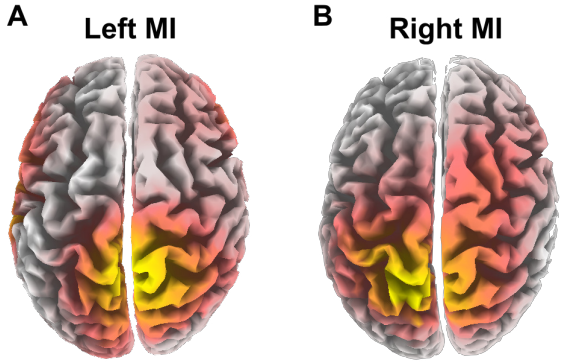

Fig. 6. Source localization plot in the beta band (12.5-30 Hz) for both Left and Right MI

### B. Classification Accuracy

Classification accuracy is a crucial metric for evaluating data quality in BCI research. High accuracy indicates the algorithm's ability to differentiate classes and reflects the quality of the data used for model training and validation. It ensures that the model identifies relevant patterns, not noise, and provides an intuitive, easily interpretable measure of dataset reliability, making it a practical indicator of a dataset's suitability for BCI applications.

Two-class MI classification after ABCD yields an average classification accuracy of 93.81% [74.81%; 98.76%] (confidence interval at 95% confidence level with the "conservative" Clopper-Pearson Binomial approximation) across the 31 subjects (63 sessions) in the EyeBCI multimodal dataset (compared to 79.29% [57.41%; 92.89%] for ICA [18] and 84.05% [62.88%; 95.31%] for ASR [19]). Notably, while 7 of the 11 excluded electrodes corresponded to task-relevant regions, identified through a meta-analysis [14], their selective removal enhanced performance. This demonstrates that faulty channels introduce more disruptive artifacts than the information they carry, and that ABCD's physiological approach preserves more signal integrity than the indiscriminate data rejection of traditional methods.

The integration of SNR topographies and classification accuracy underscores the importance of channel selection and ABCD's ability to automatically use blink patterns for detecting bad EEG channels.

## IV. DISCUSSION

### A. Physiological Signatures

The ABCD algorithm innovatively repurposes blink propagation patterns — traditionally viewed as noise to be removed — as reliable physiological signatures for channel quality assessment. Unlike conventional methods that discard blink-related components entirely, our approach leverages their high effect size and spatiotemporal consistency across subjects to create individualized quality metrics. This paradigm shift is grounded in blink propagation's electrostatic properties, which follow predictable attenuation patterns when channels function properly. By treating blinks as allies rather than artifacts, ABCD preserves their utility for both channel validation and subsequent neural analysis.

### B. Etiology of Electrode Failures

Wet EEG systems are prone to characteristic failure modes stemming from both technical and anatomical factors. One primary cause is the rapid evaporation of conductive gel, which can lead to poor electrode-scalp contact and increased impedance. This issue is exacerbated by the complex nature of EEG setup and operation, where even minor procedural errors can result in suboptimal electrode placement, as well as excessive or insufficient gel application. Additionally, sweat-induced interference or short circuits can significantly degrade signal integrity, particularly in environments with high humidity or during prolonged recording sessions. Poor cap adaptation to anatomical variations, including skull shape, head diameter, and scalp curvature, can also result in electrodes not making proper contact with the scalp. These factors collectively contribute to the prevalence of bad channels, necessitating robust detection and correction mechanisms to ensure accurate and reliable EEG data.

### C. Limitations and Future Directions

While the ABCD algorithm demonstrates robust performance in offline analysis, several considerations merit attention for real-world implementation. First, although designed for real-time operation, the algorithm's computational efficiency and responsiveness require empirical validation in true real-time BCI settings. The method's reliance on consistent blink patterns introduces two constraints: individual variations in blink frequency may affect quality assessment reliability, and task-dependent blink suppression (e.g., during intense visual fixation) could temporarily limit data availability. Furthermore, while the 62-channel wet EEG validation confirms core functionality, performance with alternative configurations—including high-density arrays (>128 channels) or dry electrode systems—remains untested, despite their growing clinical adoption.

The current validation dataset, though sufficiently powered (N=31 subjects, 63 sessions), was limited to healthy adults, leaving open questions about performance in populations with atypical ocular activity (e.g., Parkinson's disease patients with reduced blink rates or individuals with dry eye syndrome). Additionally, long-term stability assessment is needed, as factors like progressive gel drying or cap displacement during extended use may differentially impact ABCD's performance compared to shorter laboratory recordings. These findings collectively highlight the importance of future studies addressing: (1) real-time operational benchmarks, (2) population-specific adaptations, and (3) technological generalization across diverse EEG hardware paradigms.

## V. CONCLUSIONS

This study presents a novel approach to enhancing Brain-Computer Interface (BCI) performance by optimizing the signal-to-noise ratio (SNR) through the automatic detection and removal of faulty EEG channels. The Adaptive Blink-Correction and De-Drifting (ABCD) algorithm leverages blink propagation patterns to identify and eliminate compromised channels. This advancement not only enhances the reliability of BCI systems but also allows for the minimization of sample sizes, making these systems more practical and efficient for real-world applications.

The implications of this work include the potential for more user-friendly and cost-effective BCI systems, facilitating broader adoption in various fields such as medical diagnostics, rehabilitation, and human-computer interaction. Future research could explore the integration of the ABCD algorithm with other signal processing techniques and its application to different neural signals and BCI paradigms. Additionally, developing real-time implementations of the ABCD algorithm could enable dynamic channel selection, further enhancing BCI performance.

In summary, this study provides a robust method for channel selection in BCI systems, leveraging physiological characteristics of blink propagation to improve signal quality and classification accuracy.